\documentclass[aps,12pt,prl]{revtex4-2}

\usepackage{bm}

\usepackage{natbib}

\usepackage{hyperref} 
\hypersetup{
    colorlinks = true,
    linkcolor={gray},
    citecolor={blue!50!black},
    urlcolor={blue!50!black}
}
\usepackage{siunitx}
\usepackage{mathtools}
\usepackage{bbold}
\usepackage[normalem]{ulem}
\usepackage{enumerate}
\usepackage[dvipsnames]{xcolor}
\def\be{\begin{equation}}
\def\ee{\end{equation}}
\def\e#1{\label{#1}\end{equation}}
\def\bea{\begin{eqnarray}}
\def\eea{\end{eqnarray}}
\def\ea#1{\label{#1}\end{eqnarray}}

\def\bem#1{\begin{mathletters}\label{#1}}
\def\eml{\end{mathletters}}

\newcommand{\ket}[1]{\left| #1 \right>}

\def\4#1{{\boldsymbol{#1}}}
\def\8#1{{\widetilde{#1}}}
\def\bse{\begin{subequations}}
\def\ese{\end{subequations}}

\newcommand{\iu}{{i\mkern1mu}}

\begin{document}

\title{Matters Arising: Entanglement-enhanced matter-wave interferometry in a high-finesse cavity}

\author{Liam P. McGuinness}
\affiliation{Laser Physics Centre, Research School of Physics, Australian National University, Acton, Australian Capital Territory 2601, Australia \\ \textnormal{Email: \href{mailto:liam@grtoet.com}{liam@grtoet.com}} }

\begin{abstract}
In their paper ``Entanglement-enhanced matter-wave interferometry in a high-finesse cavity" Nature (2022) \cite{Greve2022}, Greve et.~al. claim to use entanglement in a matter-wave interferometer to achieve a sensitivity beyond that achievable with the same number of independent particles -- a limit known as the standard quantum limit (SQL). In particular, using squeezed momentum states of 700 atoms, the authors claim to directly observe a sensitivity 3.4\,dB (a factor of 1.5) below the SQL. This claim is incorrect. The authors do not measure anything beyond the SQL, nor do they achieve a sensitivity beyond what one could obtain with a single atom. The achieved sensitivity is at least a factor of 39 worse than the claimed value.
\end{abstract}

\maketitle

\pagestyle{plain}
In ``Entanglement-enhanced matter-wave interferometry in a high-finesse cavity" Nature (2022) \cite{Greve2022}, Greve et. al. describe measuring the phase $\phi$ between two quantum states $\ket{a}, \ket{b}$, given by $\frac{1}{\sqrt{2}}\left(\ket{a} + e^{\iu \phi}\ket{b}\right)$. With no prior information on $\phi$, i.e. $0 < \phi \leq 2\pi$, one cannot estimate this phase with a single measurement to an angular uncertainty better than $\Delta \phi = 1$\,radians. Allowing for $N$ trials, which can be implemented by encoding $\phi$ into the state of $N$ independent atoms, the uncertainty must be greater than $\Delta \phi_{\mathrm{SQL}} = 1/\sqrt{N}$\,rad, a limit known as the standard quantum limit (SQL). As Greve et. al. note, this is a fundamental limit which cannot be improved upon, even with entanglement. In fact, if one entangles $N$ atoms to obtain the state $\frac{1}{\sqrt{2}}\left(\ket{a_N} + e^{\iu \phi}\ket{b_N}\right)$, where $\ket{a_N}, \ket{b_N}$ are states in an $N$-dimensonal Hilbert space, measurement of this phase has an uncertainty is restricted to $\Delta \phi > 1$\,rad, much worse than the SQL. The reason being that we have lost the ability to perform $N$ independent trials and are thus limited to the uncertainty of a single measurement. This is clear to see, since the entangled state is identical to the single atom state up to a relabelling.

While the above is well-known for estimating an unknown quantum phase, it is widely accepted that despite this entanglement can lead to improved estimation of some other phase $\theta$. Why is this? The argument is that entangled states accumulate a bigger quantum phase in response to some physical Hamiltonian to be measured ($N$-fold greater than a single atom). With particular reference to Mach-Zehnder interferometry, discussed by Greve et. al., this is the phase that particles in one arm of the interferometer accumulate with respect to particles in the other arm. If we also allow somewhat sneakily that we now have more prior information on $\phi$ -- it is known to within a much narrow range, then it is expected that, for the same measurement time, entangled states achieve greater sensitivity to small phase shifts than unentangled states \cite{Note1}. As a result, although the uncertainty in estimating $\phi$ is $\sqrt{N}$ worse with entanglement than with independent atoms, the value of $\phi$ in an entangled state is $N$-fold greater than any of the independent atomic states. Again, it is important to be clear here. With entanglement we have not improved the uncertainty in estimating $\phi$, it has gotten worse. Whenever measurement of a quantum phase is performed as described above, one can immediately rule out surpassing the SQL, it is only when the quantum phase is used to infer some other parameter $\theta$ that the uncertainty in $\theta$ can be reduced.

Let's assume an element with phase $\theta$ is in one arm of the interferometer, and passing a single atom through the interferometer performs a one-to-one mapping of $\theta$ to the atomic phase $\phi$. Then with no entanglement we have $\phi = \theta$, whereas with entanglement $\phi = N\theta$. In the former case (assuming no additional errors) $\Delta \theta = \Delta \phi$ and estimation of $\theta$ with $N$ atoms is limited by the SQL to: $\Delta \phi > 1/\sqrt{N} \Longrightarrow \Delta \theta > 1/\sqrt{N}$. With entanglement $\Delta \theta = \Delta\left(\phi/N\right)$ again assuming no additional errors, so we obtain $\Delta \phi > 1 \Longrightarrow \Delta \theta > 1/N$. If the second inequality can be saturated with no overheads, then entanglement outperforms unentangled sensors. Contrary to what is often stated, this superiority of entanglement in sensing is not proven and only holds under strong experimental assumptions. For that reason, experimental evidence with correct analysis and complete details is critical in validating the theory \cite{Note2}.

So what is the phase shown in Fig.\,1b that Greve et. al. measure with a precision beyond the SQL? Described in section \textbf{Entangled matter-wave interferometry:} as ``A relative phase accumulates between the wave packets during a free evolution time $T_{\mathrm{evol}}$", at first reading one might be surprised to find that the cause of the relative phase shift is not explicitly defined in the main text. One reason for this oversight could be that Fig.\,1b and the description of the matter-wave interferometer is somewhat misleading. Greve et. al. do not really perform Mach-Zehnder interferometry because they do not measure a phase shift between different arms of the interferometer. So what are the authors measuring? Greve et. al. allow the atoms to fall through a gravitational potential and measure the energy shift of the atoms -- manifesting as a Doppler shift of atomic resonance with respect to the Raman laser (see Methods -- Raman transitions and velocity selection). Notably this shift is the same for both arms of the interferometer and can be measured without a Mach-Zehnder interferometer. Most importantly, entangled atoms do not experience a greater phase shift (see Fig.\,4c), since their velocity is the same as unentangled atoms.

Put simply, the authors measure the difference between the atomic phase $\phi$ and the laser phase $\theta$ at the end of the experiment. As entanglement produces no enhanced phase shift, the uncertainty in measuring either phase when using entanglement is strictly worse than the SQL, and we have shown that Greve et. al.'s claim in beating the SQL is incorrect. In fact, assuming the ensemble is fully entangled, the obtained sensitivity must be worse than the single atom precision limit. Even assuming the ensemble is not fully entangled, there are many experimental imperfections that prevent Greve et. al. reaching the single atom limit. With $N = 660$ atoms, this means that the achieved sensitivity is at least a factor of 31 worse than the claimed 1.7\,dB enhancement. Similar analysis can show that all other claims made by the authors are similarly incorrect. So how can Greve et. al. claim to have done otherwise? Maybe it is better to reframe this question with a focus on the audience. These are the questions one should ask anybody claiming to beat the SQL.
\begin{enumerate}
\item Precisely what parameter do you measure?
\item What sensitivity/uncertainty for this parameter do you explicitly achieve in your experiment? This should have the correct units, including the total measurement time and the number of particles used.
\item Is it \emph{really} impossible to measure this parameter with better sensitivity using the same measurement time and the same number of independent particles? How about just a single particle. Is it impossible to measure this parameter with better sensitivity using the same measurement time and a single particle?
\end{enumerate}

If 1) and 2) are properly defined, then the answer for all experiments to date, including the work by Greve et. al. is a resounding -- `No!'.

It is important that the scientific community is made aware of the current state of the art in quantum metrology. I am sure that many people would be extremely surprised to learn that entanglement has never been used to improve any experiment beyond what one could achieve without entanglement. Even more surprising is that fully entangled ensembles have never demonstrated a precision beyond the single particle limit. The community should be made aware of this for a variety of reasons. First, if the experimental data conflicts with the message being presented then we should demand better scientific rigour in published papers. In quantum metrology broadly, the standards have a long way to improve. Secondly, misrepresentation of data is hindering progress since people are currently unaware of a massive discrepancy between quantum mechanics as interpreted and experiment; even going so far as to prevent plausible explanations from being investigated. Thirdly, there is currently huge investment in technologies dependent on quantum entanglement (in both quantum sensing and quantum computing), if entangled ensembles cannot provide fundamentally more information than a single atom then these technologies will never reach their goals.

\section{Communication with Authors and Nature editor}
\noindent On Dec 22, 2022 a copy of this critique and a request for response was sent to the authors. No response was received. On Jan 9, 2023 a follow-up email was sent to the authors. No response was received.

\noindent On Jan 11, 2023, this critique, supplemented by correspondence with the authors, was submitted to \emph{Nature} to be published as a Matters Arising. On Feb 1, 2023 the manuscript was rejected.


\begin{thebibliography}{3}
\expandafter\ifx\csname url\endcsname\relax
  \def\url#1{\texttt{#1}}\fi
\expandafter\ifx\csname urlprefix\endcsname\relax\def\urlprefix{URL }\fi
\providecommand{\bibinfo}[2]{#2}
\providecommand{\eprint}[2][]{\url{#2}}

\bibitem{Greve2022}
\bibinfo{author}{Greve, G.~P.}, \bibinfo{author}{Luo, C.},
  \bibinfo{author}{Wu, B.} \&
  \bibinfo{author}{Thompson, J.~K.}
\newblock \bibinfo{title}{Entanglement-enhanced matter-wave interferometry in a high-finesse cavity}.
\newblock \emph{\bibinfo{journal}{Nature}} \textbf{\bibinfo{volume}{610}},
  \bibinfo{pages}{472--477} (\bibinfo{year}{2022}).
\newblock \urlprefix\url{https://doi.org/10.1038/s41586-022-05197-9}.

\bibitem{Note1}
\newblock \bibinfo{title}{I am presenting the view of the general quantum metrology community, however I must make my stance clear. I do not share the same views. Indeed I expect that entanglement cannot improve the amount of information that can be obtained from the same number of atoms in the same time. Specifically the information gain from fully entangled states cannot exceed that obtained with a single spin in the same time}.

\bibitem{Note2}
\newblock \bibinfo{title}{One specific overhead is the total measurement time $T$. For example by increasing the interaction time of a single atom with the phase element we can also achieve a mapping $\phi = N\theta$ with a single atom. This is why $\Delta \theta > 1/\sqrt{N}$ is not the SQL for estimating $\theta$. The SQL for estimating $\theta$ is $\Delta \theta > 1/(\Omega T \sqrt{N})$, where $\Omega$ is the interaction strength between the atom and the phase element.}


%
%
%
%
%
%
%
%
%
%
%
%
%
%
%
%
%

\end{thebibliography}
\end{document}